\documentclass{article}

\usepackage{amsmath, amsthm, amssymb}
\usepackage{algorithm}
\usepackage[noend]{algorithmic}
\newtheorem{theorem}{Theorem}
\newtheorem{proposition}{Proposition}
\theoremstyle{definition}
\newtheorem{algo}{Algorithm}

\usepackage{arxiv}

\usepackage[utf8]{inputenc} 
\usepackage[T1]{fontenc}    
\usepackage{hyperref}       
\usepackage{url}            
\usepackage{booktabs}       
\usepackage{amsfonts}       
\usepackage{nicefrac}       
\usepackage{microtype}      
\usepackage{cleveref}       
\usepackage{lipsum}         
\usepackage{graphicx}
\usepackage{natbib}
\usepackage{doi}

\title{Practical Fixed-Parameter Algorithms for Defending Active Directory Style Attack Graphs}
\author{Mingyu Guo, Jialiang Li, Aneta Neumann, Frank Neumann, Hung Nguyen
\\
    School of Computer Science\\
    University of Adelaide, Australia\\
    \{mingyu.guo, j.li, aneta.neumann, frank.neumann, hung.nguyen\}@adelaide.edu.au
}

\begin{document}

\maketitle

\begin{abstract} Active Directory is the default security management system for
    Windows domain networks.  We study the shortest path edge interdiction
    problem for defending Active Directory style attack graphs.  The problem is
    formulated as a Stackelberg game between one defender and one attacker.
    The attack graph contains one destination node and multiple entry nodes.
    The attacker's entry node is chosen by nature.  The defender chooses to
    block a set of edges limited by his budget.  The attacker then picks the
    shortest unblocked attack path.  The defender aims to maximize the expected
    shortest path length for the attacker, where the expectation is taken over
    entry nodes.

We observe that practical Active Directory attack graphs have small maximum attack
    path lengths and are structurally close to trees.  We first show that even
    if the maximum attack path length is a constant, the problem is still $W[1]$-hard
    with respect to the defender's budget.  Having a small maximum attack path length
    and a small budget is not enough to design fixed-parameter algorithms.  If
    we further assume that the number of entry nodes is small, then we derive a
    fixed-parameter tractable algorithm.

We then propose two other fixed-parameter algorithms by exploiting the
    tree-like features. One is based on tree decomposition and requires a small
    tree width. The other assumes a small number of splitting nodes (nodes with
    multiple out-going edges).  Finally,  the last algorithm is converted into
    a graph convolutional neural network based heuristic, which scales to
    larger graphs with more splitting nodes.
\end{abstract}

\section{Introduction}

Cyber attack graphs model the chain of events (conceptual or physical) that
lead to successful cyber attacks.  Despite its popularity in both academia and
industry, there is not a {\em canonical} definition of cyber attack graphs.
\citet{Lallie2020:reviewa} surveyed over $180$ attack graphs/trees studied in
literature, and discovered over $90$ different {\em self-nominated} definitions
of attack graphs/trees.

For industry practitioners, there is one attack graph model that stands out and
finds its place in many practitioners' toolkit, which is the Active Directory
attack graph.  Microsoft Active Directory is the {\em default} security
management system for Windows domain networks, which has a dominant market share
among large organisations worldwide.  Due to its popularity, Active Directory
has been a focused cyber attack target.  An Active Directory environment
naturally describes an attack graph, where the nodes represent accounts,
computers, security groups, etc.  A directed edge from node A to B represents
that an attacker can reach from A to B via existing accesses or known exploits.
There are a number of software tools for analysing/visualising Active Directory
attack graphs.
Among these tools, {\sc BloodHound}\footnote{https://github.com/BloodHoundAD/BloodHound}
is the most influential.  Motivated by
\cite{Dunagan2009:Heatray}, {\sc BloodHound} models the {\em identity snowball
attack} under Active Directory.  Typically, an identity snowball attack starts
when an attacker gains initial access to the internal
network. The attacker starts from a low-privilege user account (often
obtained via phishing emails).  The attacker then moves from low-privilege
nodes to high-privilege nodes ({\em i.e.}, account A
$\xrightarrow[\text{AdminTo}]{\text{admin access}}$ computer B
$\xrightarrow[\text{HasSession}]{\text{scan memory}}$ account C).  The goal of
the attacker is to reach the highest-privilege account, called the {\em Domain
Admin {\sc DA}}.  The core functionality of {\sc BloodHound} is to automatically generate the
{\em shortest attack path} from the attacker's entry node to {\sc DA},
where the distance is defined as the number of {\em hops}.
Following the shortest attack path implies less time spent on the attack
and less chance of failure. Before the invention of {\sc BloodHound},
attackers used to explore aimlessly in the internal network hoping to discover
a privilege escalation pathway.
\citet{Dunagan2009:Heatray} briefly described a
heuristic edge blocking algorithm. The aim is to block a small number of edges
to cut the attack graph into multiple disconnected regions, which essentially
removes the attack paths from most entry nodes.

We derive {\bf optimal} edge blocking policies. The defender can assign different utilities
on attack paths of different lengths. That is, maximizing the number of attack paths
cut is a special case of our model.
We adopt a two-player {\em Bayesian
Stackelberg game}~\cite{Paruchuri2008:Efficient} setup with {\em pure strategies only}.  In our game, the defender
(leader) has a limited {\bf budget} $b$ for blocking edges. Not all edges are
blockable.  The attacker's type is characterized by his entry node.  In
practise, the entry account is often from a phishing attack victim. Therefore,
we assume that the attacker's entry node is drawn randomly by nature from
a set of entry nodes ({\em i.e.}, users whose emails are listed on the organisation's website).  The
attacker is the follower in the Stackelberg game.  That is, the attacker is
aware of which edges have been blocked.\footnote{In practise, the attacker
can use a tool called {\sc SharpHound} to scan the environment to obtain
information on all edges.} We assume that the attacker follows {\sc BloodHound}'s
advice and attacks via a shortest attack path.  The defender aims to
maximize the attacker's expected shortest path length, where expectation is
taken over all entry nodes.

When there is only one entry node, our model reduces to the well-studied
shortest path {\em edge interdiction} problem (also called the {\em most vital
edges} problem). \citet{Bar-noy1995:Complexity} already showed that the problem
is NP-hard.
Nevertheless, this negative result does not rule out scalable algorithms for
{\em practical} Active Directory attack graphs.  We adopt {\em parameterized
complexity analysis} and design {\em fixed-parameter tractable (FPT)}
algorithms.  Fixed-parameter algorithms allow us to solve some NP-hard problem
{\bf instances} {\em efficiently} and {\em optimally}, under the assumption
that the problem instances are characterized by a few parameters that are
small.  For example, vertex cover is known to be NP-complete, but the question
``whether there exists a vertex cover of size $k$'' can be solved in
$O(1.2738^k+kn)$, where $n$ is the number of nodes~\cite{Chen2006:Improved}.
That is, if one's goal is to solve vertex cover, and in one's practical
scenario $k$ is never too large, then the NP-completeness of vertex cover is
irrelevant.  Formally speaking, given problem instances with $c$ parameters
$k_1,k_2,\ldots,k_c$, the computational problem is fixed-parameter tractable if
we can design an algorithm with complexity $O(f(k_1,\ldots,k_c)poly(n))$.
That is, the complexity is allowed to be exponential in the parameters, but it
needs to be polynomial in the input size $n$.  This allows us to scale
to large input sizes, as long as the parameters are small.

\begin{figure}[h]
\begin{center}
  \includegraphics[width=0.5\linewidth]{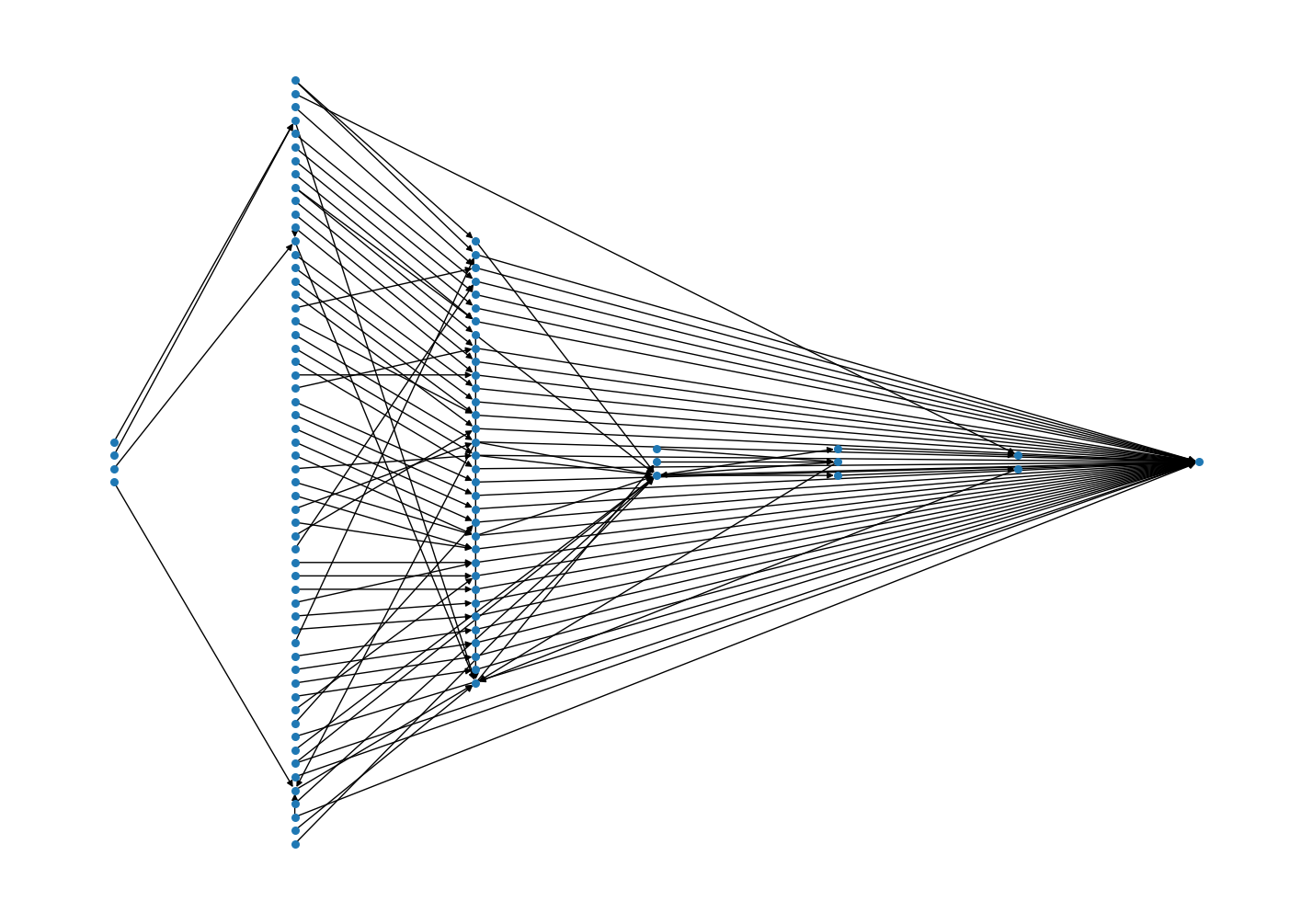}
\end{center}
    \caption{Synthetic attack graph generated using {\sc DBCreator} ($500$ computers). Only nodes reachable to {\sc DA} (rightmost node) are shown.}
  \label{fig:r500}
\end{figure}

A natural question to ask is what should be the appropriate parameters for
describing Active Directory attack graph instances. To answer this,
let us consider the following attack graph.  Figure~\ref{fig:r500} is a synthetic
attack graph generated using {\sc BloodHound} team's synthetic database generator
{\sc DBCreator}.\footnote{It should be noted that an organisation's Active
Directory attack graph is considered sensitive information. Our paper only
references synthetic graphs generated using {\sc DBCreator}. {\sc DBCreator}
generates a lot of details, such as a node's operating system, department, real names.
We only extract the topology.
} We make two observations.
First, the maximum attack path length is small.
Second, the graph looks like a tree.
Our
observation is not an artifact based upon DBCreator.  It is considered a best
practise for Active Directory to follow the organisation chart.
For example, marketing and human resources tend to form two separated tree
branches. The graph will not be exactly a tree because some computers from
human resources may need to access data from marketing. That is, an Active
Directory attack graph can be thought of as a {\em tree with additional security exceptions}
(non-tree edges connecting tree nodes to non-parents, also called the {\em feedback edges}).  The maximum attack path length
tends to be small for even the largest organisations. {\em I.e.}, it most likely only takes a few hops to
go from an intern's account to the CEO's account. This is similar to the ``six degree of separation'' idea: all people are six or fewer social connections from each other~\cite{Milgram1967:SmallWorld}.

We use $n$ and $m$ to denote the number of nodes and the number of edges.  We
adopt the following list of parameters. In experiments, we show that some of
these parameters stay tiny and our largest experiment involves
graphs with $5997$ nodes (among which $2000$ are computers).
That is, our algorithms scale well and are able to handle large organisations.

\begin{footnotesize}
\begin{center}
\begin{tabular}{ |l|l| }
 \hline
    $l$ & maximum attack path length\\
 \hline
    $w$ & {\em tree width}\\
 \hline
    $h$ & number of {\em feedback edges}\\
 \hline
    $b$ & defensive budget\\
 \hline
    $s$ & number of entry nodes\\
 \hline
\end{tabular}
\end{center}
\end{footnotesize}

Tree width is a standard measure in graph theory for describing how close a
graph is to a tree.
The optimal tree width
is NP-hard to
compute~\cite{Arnborg1987:Complexity}.
In this paper, tree width refers to the achieved tree width using our own heuristic.
A connected graph can be interpreted as a tree with
$h$ additional feedback edges. Note that $h=m-(n-1)$ as a tree has $n-1$ edges.

We first show that having a small $l$ does not make the problem easy.  We prove
that our model is $W[1]$-hard with respect to $b$ even if $l$ is a constant.
We then propose three different fixed-parameter tractable algorithms.
Our algorithms' complexities are summarized in the table below.  It should be
noted that these complexities only describe the {\em worst-case} running time.
When it comes to specific problem instances, the running time
often is significantly faster than what the table suggests.
We do not use $m$ in our complexity notation because
we assume our graphs are similar to trees.

\begin{footnotesize}
\begin{center}
    \begin{tabular}{ |l|l|l| }
 \hline
        {\sc BudgetFPT} & $O\left(l^b {b+s-1 \choose b} n\right)$ & requires small $l,b,s$\\
 \hline
        {\sc DP} & $O((l+2)^{w+1}b^2n)$ & requires small $l,w$\\
        & & acyclic graphs only\\
 \hline
        {\sc SplitFPT} & $O(3^{h}(h^2+hl+bn))$ & requires small $h$\\
 \hline
\end{tabular}
\end{center}
\end{footnotesize}





Finally, we convert {\sc SplitFPT} into a graph convolutional neural network
based heuristic. {\sc SplitFPT} enumerates which route the attacker would choose
at splitting nodes (nodes with multiple out-going edges) when facing the {\em optimal} defence.
When $h$ is too large, we cannot afford to
enumerate all scenarios.  Instead, we use a graph convolutional neural
network (GCN) to estimate the attacker's decision. Our neural network is not trained
based on real attacker data. Instead, our network is purely used as
an optimisation tool.

{\em In summary, we propose $3$ fixed-parameter algorithms that scale
in practise and a GCN-based approach. These will
help IT admins to identify high-risk edges (accesses/exploits) in practical Active Directory environments.}

\section{Related Works}

\citet{Bazgan2019:More} studied single-source single-destination shortest
path edge interdiction. The authors studied a long range of parameters.
Unfortunately, most of the parameters are irrelevant to practical Active Directory
attack graphs ({\em i.e.}, distance to clique). Nevertheless, one of the
parameters considered is the number of feedback edges, which is
also used by our {\sc SplitFPT} algorithm.  The authors proposed a
kernelization technique that converts an arbitrary graph into a graph with $6h$
edges, and then exhaustively search over all $2^{6h}$ combinations of edges.

Another similar but different problem is the bounded length cut problem, which
studies how to block $b$ edges in order to ensure that the shortest path is
greater than a parameter $l'$.  \citet{Golovach2011:Paths} proposed an elegant
FPT algorithm for solving the single-source single-destination bounded length
cut problem.  Our {\sc BudgetFPT} builds upon the core idea behind the authors'
algorithm. The authors showed that if the maximum path length and the budget
are both small, then the problem is fixed-parameter tractable. We
prove that this is not the case for our model.
\citet{Dvorak2018:Parameterized}
also studied
bounded length cut.  The authors proposed a FPT algorithm in tree width $w$ and maximum cut
length $l'$ with complexity $O(l'^{12w^2}n)$ for the single-source
single-destination model.




There have been existing works on Stackelberg games on attack
graphs~\cite{Aziz2018:Defendera,Aziz2017:Weakening,
Durkota2019:Hardening,Milani2020:Harnessinga}.
Besides models being different, all the above only discuss tiny attack graphs
with at most $50$ nodes.
Our setting and approaches are different and we deal with realistic
Active Directory attack graphs with thousands of nodes.
\citet{Wang2019:Using} applied graph neural
networks to network interdiction games.  The authors trained GCN using real
attacker data to simulate Markov chain based boundedly rational attackers.  We use graph neural networks purely as an optimization tool.
Graph neural networks have been shown to be effective for combinatorial graph
problems~\cite{Dai2017:Learning}.

\section{Model Description}

\begin{figure}[h]
    \begin{center}
  \includegraphics[width=0.5\linewidth]{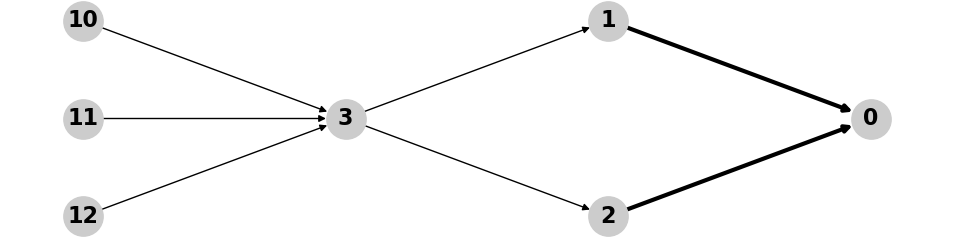}
    \end{center}
    \caption{Example attack graph. {\sc DA} is the rightmost node.}
  \label{fig:example}
\end{figure}

An Active Directory attack graph is denoted by $G(V,E)$. There are $n=|V|$
nodes and $m=|E|$ directed edges.  There is one attacker who enters the graph
via one of the {\em entry nodes}. Let $s$ be the total number of entry nodes.
We assume that the entry node is selected by nature.
For simplicity, we assume uniform chances for all entry nodes. The
attacker's goal is to reach a single {\em destination} node called Domain Admin or {\sc DA}.  We
consider only one destination.\footnote{In real-life Active Directory attack
graphs, there are often multiple admin nodes.
We simply merge all admin nodes into a single node and call it our {\sc DA}.} There is a set of edges that are
{\em blockable}, denoted by $E_b\subseteq E$. Figure~\ref{fig:example} is an example
graph. The entry nodes are $10$, $11$, and $12$. Node $0$ is the destination.
The thin edges are blockable (only $1\rightarrow 0$ and $2\rightarrow 0$ are not
blockable).

The defender selects the best $b$ edges to block, where $b$ is the defensive
budget. We use $B$ to denote the set of edges blocked.  We
consider only pure actions. That is, we do not study mixed-strategy
probabilistic blocking.  The graph after blocking is denoted as $G-B$.
We assume the attacker can
observe the defensive action and then perform a best-response attack. In the
context of Active Directory attack graphs, the attacker performs the {\em
shortest path attack} on $G-B$.  We use $SP(s_i,G-B)$ to denote the shortest
path from entry node $s_i$ to the destination, on graph $G-B$.

We assume that an attack path with $x$ hops has a success probability of
$f(x)$, where $f$ could be any decreasing function in $x$.  If no attack paths exist,
then $f(+\infty)=0$. $f$ can essentially be any performance evaluation function.
For presentation purposes, we call it the ``success rate'',
and we set $f(x)=0.95^x$ ({\em i.e.}, every action has a $5\%$ failure rate).

The defender's task is to pick the best $b$ edges to block, in order to
minimize the attacker's expected success rate. Expectation is taken by
averaging over all $s$ entry nodes. Formally, our model is the following:

\begin{equation*}
\begin{aligned}
    \min_{B\subseteq E_b,|B|\le b} \quad & \frac{1}{s}\sum_{i=1}^sf(|SP(s_i,G-B)|)\\
\end{aligned}
\end{equation*}

A baseline algorithm is the greedy algorithm ({\sc Greedy}).  We present it
here to serve as an example to help readers understand our model.
The greedy algorithm will also be used as an algorithm component in later sections.

\begin{algo}[{\sc Greedy}] Given budget $b$, we pick the single best edge to
    block in terms of reducing the attacker's expected success rate.\footnote{For
    example, we could use the almost linear algorithm proposed in
\cite{Nardelli2001:faster} for each step.}  We repeat $b$ times.  \end{algo}

{\sc Greedy} is actually the optimal algorithm if our graph is exactly a tree.
This is straightforward to see. Blocking actions happening in different tree branches
are independent from each other. Given edge $e_1$ and $e_2$, if the path from $e_1$
to {\sc DA} passes through $e_2$, then {\sc Greedy} will always block $e_2$.


On the other hand, {\sc Greedy} works poorly if there are {\em substitutable block-worthy} edges.
For Figure~\ref{fig:example}, if our budget is $2$, under {\sc Greedy}, the edges
blocked are any two of $\{10\rightarrow 3,11\rightarrow 3, 12\rightarrow 3\}$.
Neither $3\rightarrow 1$ nor $3\rightarrow 2$ is blocked.  There is still one entry node that can reach {\sc DA}. The attacker's
success rate is $\frac{1}{3}f(3)$.  It is easy to see that the optimal defence
should be blocking both $3\rightarrow 1$ and $3\rightarrow 2$.


\section{$W[1]$ Hardness and {\sc BudgetFPT}}

We mentioned that Active Directory attack graphs tend to have short
attack paths.  We first prove that having a small maximum attack path length alone is not
enough to derive efficient algorithms.

\begin{theorem}
    \label{thm:w1}
    Let $b$ be the budget.
    Let $l$ be the maximum attack path length.
    Our problem is $W[1]$-hard with respect to $b$ even for constant $l$.
\end{theorem}


Theorem~\ref{thm:w1} implies that having a small maximum attack path length and a
small budget is not enough to derive fixed-parameter algorithms, which makes it
different from the single source single destination bounded length cut problem
studied in \cite{Golovach2011:Paths}.
Nevertheless, the authors' core idea
still applies to our setting.  The core idea goes like this: For single
source and single destination, we pick an arbitrary shortest path. We must
block somewhere along this path. Otherwise, it is as if no blocking happens.
Our fixed-parameter algorithm builds on the above idea.




\begin{algo}[{\sc BudgetFPT}]
We go over all combinations to find the blocking setup that minimizes the expected success rate for the attacker.
    We pick an arbitrary entry node $s_1$ and pick an
    arbitrary shortest path from $s_1$ to {\sc DA}. We either block at least
    one edge along this shortest path, or $s_1$ can be removed from our
    consideration (as its shortest path is unaffected). There are at most $l$
    options for picking an edge to block.  After blocking an edge, the budget
    is reduced by $1$.  There is an option for ignoring $s_1$. After
    ignoring $s_1$, the number of entry nodes is reduced by $1$.

We use $c_{b,s}$ to denote the number of combinations we need to
go over when the remaining budget is $b$ and the remaining number of entry
nodes is $s$.
We have the following recursive relationship and the base cases:
\[c_{b,s} = lc_{b-1,s} + c_{b,s-1};\quad c_{0,s} = 1;\quad c_{b,1} = l^b \]

\end{algo}

For each combination, the success rate evaluation takes linear time.
We derive the analytical form of $c_{b,s}$, which gives the following
complexity result.

\begin{proposition}
    \label{prop:budget}
    {\sc BudgetFPT} has a complexity of
    \[O\left(l^b {b+s-1 \choose b} n\right)\]
\end{proposition}



\section{Tree Decomposition based Dynamic Program}

In this section, we propose a dynamic programming algorithm based on tree decomposition.
This algorithm scales the best in experiments, but one restriction is that it
requires the attack graph be acyclic.  The synthetic attack graph in
Figure~\ref{fig:r500} does contain cycles.  However, all cycles
involve one common user. So if we remove that one user, the graph becomes
acyclic. In other graphs generated using {\sc DBCreator}, we always only need
to remove a few nodes to make the graph acyclic.  We argue that the
network admin has the flexibility to convert existing graphs into
acyclic with minimum changes. Our algorithm can also serve as a heuristic.

The {\em tree width} of a graph is a
commonly used index to measure how close a graph is to a tree. Many
computationally difficult problems on general graphs are easy if the tree width
is small. This is also the case for our problem.










Our attack graphs are
directed and all nodes have paths to {\sc DA} (nodes that cannot reach {\sc DA}
are ignored).  We further assume that the attack graphs are acyclic in this
section.  Under the above assumptions, we could use topological sorting to
divide our attack graphs into several {\em security levels}. {\sc DA} belongs
to the highest security level. An edge only goes from a lower security level node to a
{\em strictly} higher security level node.  Take Figure~\ref{fig:example} as an example, we could
interpret it as that there are four security levels. $10$ to $12$ are in the
lowest security level while {\sc DA} (node $0$) has the highest security level.
With this security-level interpretation, there is no ambiguity on the direction
of an edge. If an edge involves node $a$ and $b$, then the direction must be
from $a$ to $b$ if $a$ has a lower security level.  We will directly reuse the
tree decomposition and tree width definition for undirected graphs.

\begin{figure}[h]
    \begin{center}
  \includegraphics[width=0.5\linewidth]{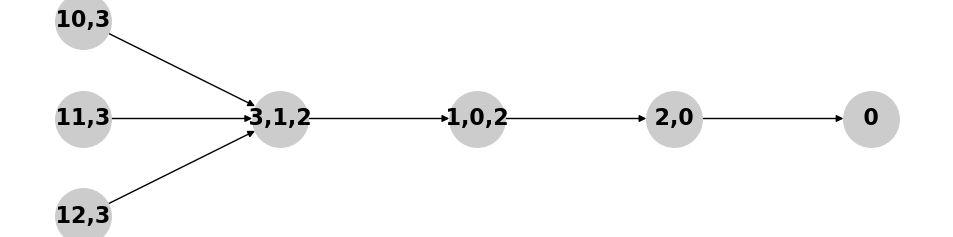}
    \end{center}
    \caption{Tree decomposition of Figure~\ref{fig:example} using our heuristic}
  \label{fig:example_td}
\end{figure}

As an example, Figure~\ref{fig:example_td} shows a tree decomposition of the
graph in Figure~\ref{fig:example}. Every tree node contains a subset of the
original graph vertices. The union of all tree nodes contains all graph
vertices.  Every edge is covered by a tree node. For example, tree node
$(3,1,2)$ covers all out-going edges of graph vertex $3$.  Finally, for every
graph vertex, the tree nodes containing it form a subtree.
For example, the tree nodes containing $3$ are $(3,1,2)$, $(10,3)$, $(11,3)$
and $(12,3)$.

Tree decompositions are not unique. The optimal tree decomposition that
minimizes the tree width is NP-hard to
compute~\cite{Arnborg1987:Complexity}.  For our dynamic program, we require a
specific style of tree decompositions that satisfy extra properties.  We call
such tree decompositions {\em desired} tree decompositions.
Figure~\ref{fig:example_td} is a desired tree decomposition.  We first explain
how we can convert a desired tree decomposition into a dynamic program via an example. During
the process, we will explain what it means for a tree decomposition to be
desired. Finally, we propose a heuristic that guarantees to generate a
desired tree decomposition.

In Figure~\ref{fig:example_td}, the number of tree nodes is the same as the
number of graph vertices. The first coordinate values are unique.  A dynamic
program subproblem involves a tree node and a remaining budget $b'$.  For
example, $((3,1,2),b')$ corresponds to the following subproblem: given the
remaining budget $b'$ and the distances (to {\sc DA}) for $1$ and $2$, what
should be the distance between $3$ and {\sc DA}?  To answer this question, we
essentially need to determine the budget investment on $3$ (budget spent on $3$
is used to block $3$'s out-going edges). $3$ has two successors ($1$
and $2$). Both successors are blockable, so we could spend $0$ to $2$ units
of budget on $3$. If $x$ units of budget is spent on a node (to block its out-going edges), it will always be blocking the shortest $x$ paths.  Once the budget investment on $3$ is determined, $3$'s final
distance is also determined.  To summarize, $((x,y_1,y_2,\ldots),b')$ is the
subproblem where the defender needs to decide how many units of budget (at most $b'$) are invested on $x$, considering the {\em context information}
on $x$'s potential successors (the $y_i$).
Here, context information refers to the distances to {\sc DA}.

Let us run through a solution on the basis of Figure~\ref{fig:example_td}.  Let
us recall that the actual graph is in Figure~\ref{fig:example}.  We still
assume the total budget is $2$.  We start from $((0),2)$. The budget investment
on $0$ will be $0$ since its distance to itself is always $0$. This information
is then passed down to $((2,0),2)$.  $2$ has no blockable out-going edges, so
we cannot spend any budget anyway. The distance (to {\sc DA}) for $2$ is then
$1$.  This information is passed down to $((1,0,2),2)$.  Again, we cannot
invest any budget because $1$'s out-going edge is not blockable. We get that
$1$'s distance is $1$. This context is passed down to $((3,1,2),2)$.  At this
point, we are deciding how many units of budget to spend on $3$.  Based on the
context information, we know that $1$ and $2$'s distances are both $1$.  The
best decision here is to invest $2$ units of budget. After that, $3$'s distance
becomes infinity. This information is then passed down to all three leaf nodes.
For example, one subproblem at the leaf level is $((10,3),0)$. The context
information says that $10$'s successor $3$'s distance to {\sc DA} is infinity,
and there is no budget left. So we get $10$'s distance to {\sc DA} is also
infinity. Since $10$ is one of three entry nodes, we add
$\frac{1}{3}f(+\infty)=0$ to the expected success rate. Essentially, the
original problem is $((0),b)$. Our dynamic program makes decisions on the budgets spent on
each node. Context information and remaining budgets are passed down to future
subproblems.  Through out the process, one property we need is that as we move
from the root to the leaves, when a vertex first enters into our consideration
({\em i.e.}, vertex $3$ enters at tree node $(3,1,2)$), the context information
({\em i.e.}, $1$ and $2$'s distances) must be enough for us to decide the new
vertex' distance right away.  That is, for any vertex $i$, we consider the subtree
containing $i$.
The root of this subtree must contain all out-going
edges of $i$.  Also, the root of the whole tree should be a node containing
{\sc DA} only.  The above properties are not held by all tree decompositions.
In this paper, we use the vertex elimination technique to generate tree
decomposition~\cite{Bodlaender2006:Exact}. This technique maps an arbitrary
permutation of the vertices into a tree decomposition. It should be noted that
the vertex elimination technique is only a heuristic
framework because it is NP-hard to figure out the best permutation that results
in the smallest tree width. Our heuristic is simple: eliminate vertices
based on the security level ranking, with the lowest security level eliminated first.
We of course
cannot guarantee that the generated tree width is the smallest, but it is
provable that it generates a tree decomposition that is desired.

Lastly, actually our example in Figure~\ref{fig:example_td} does not capture
one complication, which is that at node $(3,1,2)$, we also need to divide the remaining budget among
three child branches. We need to divide it four ways (one for $(3,1,2)$
itself, for blocking $3$'s out-going edges; and three for three child branches).
This has a complexity of $b^4$. If there are too many branches, then the decision
process gets too expensive. We resolve this using the {\em nice tree
decomposition} idea from \cite{Cygan2015:Parameterized}. The main idea is that
we can always clone a node into two. For example, we can insert another
$(3,1,2)$ in between $(3,1,2)$ and $(1,0,2)$.  The copy closer to {\sc DA} only
makes the decision on $3$ itself (how many units of budget to spend on $3$).
The copy further away from {\sc DA} deals with the splitting problem for splitting
the remaining budget for three children. However, this node still needs to split three ways.
The nice tree decomposition idea can resolve this with ease as well. We simply need
to add another clone of $(3,1,2)$ to handle the second round of splitting.
One node splits between $(10,3)$ and the rest $\{(11,3), (12,3)\}$.
The second node splits between $(11,3)$ and $(12,3)$.
For our setting, the number of clone nodes needed is at most $2n$, which
does not affect the complexity. The gain is that for every node, the decision
is always one dimensional, so the number of decisions at a node is at most $b+1$.

\begin{proposition}
    \label{prop:dp}
    {\sc DP} has a complexity of
    \[O((l+2)^{w+1}b^2n)\]
    $w$ is the {\em achieved} tree width using our desired tree
    decomposition heuristic.
\end{proposition}

\section{Classification-based {\sc SplitFPT}}

Another parameter to describe a graph's tree-likeness is the number of feedback
edges $h$.
A related parameter is the number of splitting nodes $t$ (nodes with multiple
out-going edges). If the maximum out degree is $d$, then we have $h \le t(d-1)$.
For the synthetic graph in Figure~\ref{fig:r500}, the number of splitting nodes
is only $12$, and the maximum out degree $d$ is only $3$.
Having small $t$ and $d$ essentially means that the attacker's strategy space
is tiny.  If the attacker starts from a non-splitting node, then the attacker
has no choice but to move on to the only successor, and keeps moving on until 1) the
attacker is facing a blocked edge, which means the attacker's entry node cannot reach {\sc
DA}; 2) the attacker has reached {\sc DA}; or finally 3) the attacker has reached another
splitting node. A path where every intermediate node has only one out-going edge
is called a {\em simple path}. At a splitting node, an attacker faces at most $d$
simple paths (one for each successor). Every simple path leads to either
another splitting node or to {\sc DA}. Essentially, the attacker's strategy is
characterized by his route choices at the splitting nodes.  At each splitting
node, there are at most $d+1$ options, including at most $d$ simple paths to choose from, and not choosing at all, which happens if none of the simple paths leads to {\sc DA}.
The attacker's strategy space has a size of $(d+1)^t$.

Typically for Stackelberg games, we optimize over the defender's strategy
space.  For each defender's strategy, we figure out the attacker's
best-response, and then check how good this best-response attack
is in terms of the defender's utility.
In this paper, because the attacker's
strategy space is tiny, we instead {\em optimize over the attacker's strategy
space}.
We {\bf guess} what the attacker would do when facing the {\bf optimal}
defence.  That is, we guess the attacker's route choices at the splitting nodes.
When both $t$ and $d$ are tiny, we can simply go over all strategies of the
attacker. If we go over all combinations, then at least one is
indeed the best response to the optimal defence.
Given a guessed attacker's strategy, we can derive the distances (to {\sc DA})
from all splitting nodes. For example, given split node $t_1$, we know what
route the attacker would choose at this node, which has to be the shortest unblocked route,
since we assume it is a best response.
We can follow along this
simple path to another splitting node (or to {\sc DA}). At the next splitting
node, we also know what route the attacker would choose. This way we can generate the
shortest paths from all splitting nodes to {\sc DA}, except for splitting
nodes at which the attacker chooses not to take any route (the distances are
set to infinity for these nodes).  For all generated
shortest paths, we mark all blockable edges covered by them as {\em not blockable}.
The defender cannot block these edges, otherwise it is contradictory to
our assumption that the attacker's strategy is a best response.
Then, at each splitting node,
there are routes not taken by the attacker (either all but one are not taken,
or simply all are not taken).  Let us consider a splitting node $a$, and one of
its successors $b$, where route $a\rightarrow b$ is not taken. We follow the simple path starts from $a\rightarrow b$, and continue on until we either reach {\sc DA} or another splitting node.
We always end up with a node whose distance to {\sc DA} is known.
We calculate whether the simple path under discussion is a better choice for the attacker or not. If not, we do not need to do anything.
If it is a better choice for the attacker, then we must block this simple path.
Again, otherwise it is contradictory to our assumption that the attacker's strategy
is the best response to the optimal defence.
For a simple path, it
is without loss of generality to block the blockable edge that is closest to {\sc DA}.
There is no reason to block further away edges or multiple edges.  After the above
actions (marking some edges as not blockable and blocking some edges), we end
up with an attack graph that is exactly a tree (only one rational decision
at any splitting node).
When the attack graph is a tree, we simply
run {\sc Greedy} using the remaining budget.
\begin{proposition}
    \label{prop:split}
    {\sc SplitFPT} has a complexity of
        \[ O(3^{h}(h^2+hl+bn))\]
\end{proposition}

\section{Graph Convolutional Neural Networks}

{\sc SplitFPT} exhaustively goes over the attacker's $d+1$ options at all $t$
splitting nodes. This is exactly a {\em node classification} problem.  We use graph
convolutional neural network to perform node classification so that it
scales to larger graphs.

We perform unsupervised learning as follows.  A node classification neural
network takes one splitting node as input, and outputs its classification ({\em
i.e.}, which route the attacker would take when facing the optimal defence; the
attacker may take no routes, which is just another classification category).
Given a classification on the splitting nodes, we can evaluate the
corresponding defence that would induce this classification using the same
procedure in {\sc SplitFPT}.  That is, we can map a node classification to an
expected success rate for the attacker.

A short description of our neural network approach is that, given the current
neural network (the current node classification rule), we randomly flip some nodes'
classifications to check whether the perturbed classification leads to better
result (worse success rate for the attacker). If so, we instruct the neural
network to learn toward the better classification (treat it as {\em true labels}).  We obviously
do not have the computational resources to exhaustively go over all node
classification combinations (otherwise we should simply run {\sc SplitFPT}). The goal is to
get a small number of nodes' classifications correct, and hope the neural network is able
to pick up the underlying rules, and can help generalize to produce
correct classifications on all nodes.

Below are the details of our neural network.

\noindent{\bf Node and edge features:} Because the original graph is too large
to handle, we construct a {\em condensed graph}
containing only the splitting nodes.  The splitting nodes are connected via
simple paths in the original graph. In the condensed graph, a simple path is
interpreted as a single edge. An example node feature is its {\em out degree}. An
example edge feature is {\em the length of the corresponding simple path}.
Due to space constraint, we omit our list of manually derived features ($7$ node
features, $6$ edge features).

\noindent{\bf Network structure:} Both the node features and the edge features
are expanded to $64$ dimensions using linear encoders. The features then go
through $10$ layers of {\em crystal graph convolutional layer} with {\sc max}
as the aggregator and batch normalization turned on~\cite{Xie2018:Crystal}.
Between each layer, we have a dropout layer that drops an edge with $0.1$
probability.  The last layer is a linear layer that converts the output to
$d+1$ dimensions. Given an output, the coordinate with the highest value is
taken as the neural network's classification for the input splitting node.

\noindent{\bf Training:} We use a batch size of $16$ (splitting nodes). For
splitting nodes not in the batch, their classifications follow the current
network's decision.  For a node in the batch, with $0.9$ probability, we follow
the current network's classification, but we do not just pick the coordinate
with the highest value. Instead, we use {\sc softmax} to get each dimension's
probability and we draw a classification accordingly.  With $0.1$ probability,
we disregard the network and draw a classification uniformly at random.
Essentially, we slightly perturb the in-batch nodes' classifications.  We then
go over the in-batch nodes one by one based on a random permutation order. For
each node, we exhaustively go over all $d+1$ classifications and check whether
unilateral change leads to a better-performing classification.  We end up with
a new classification. With $0.5$ probability, we use this new classification as
true labels.  With the other $0.5$ chance, we use the historically
best-performing classification as true labels.  We stop after $50$ epochs.
Loss is based on cross entropy and the optimizer is Adam with a learning rate
of $0.01$.  We also always train $5$ times with random seed $0$ to $4$.

\section{Experiments}

Our hardware specs are {\sc i7-6700 3.4GHz} and double {\sc TURBO-GTX1080-8G} GPUs.
We conduct all experiments using a synthetic Active Directory attack graph
generated by {\sc DBCreator}. We set the number of computers to $2000$.  We
only consider three types of edges: {\sc AdminTo}, {\sc MemberOf}, {\sc
HasSession}. These three edge types are the only three {\em default} edge types
in {\sc BloodHound}.  The final graph contains $5997$ nodes (computers + user
accounts + security groups, etc.) and $18795$ edges.\footnote{
There are $7$ admin
nodes. We merge them into one node {\sc DA}.
Out of $5997$ nodes, only $339$ can reach {\sc DA}.
So we can preprocess our graphs to make it much smaller.
}
We call this graph {\sc r2000}.  We can make {\sc
r2000} acyclic by removing another $4$ nodes from it. We call the acyclic
version {\sc r2000-dag}.

We randomly set some edges to be blockable. For a non-splitting node, its
out-going edge is blockable with $p_{b}$ probability. For a splitting node, with
$p_{b}$ probability, all of its out-going edges are blockable.
We set $p_{b}$ to be $0.2$.
All of our experiments are repeated $10$
times.

\noindent {\bf Table~\ref{tab:experiment1}: Low budget setting.}
We randomly pick
$5$ nodes to be the entry nodes.
We set the budget to $10$.
With a small budget and a small number of entry nodes, {\sc BudgetFPT} is scalable.
We compare {\sc BudgetFPT} against {\sc Greedy} and {\sc GCN} (the graph convolutional
neural network approach). Note that {\sc BudgetFPT} is optimal.
Success Rate is short for the expected success rate for the attacker (lower is better).
Time[s] is seconds per trial. \#Opt shows how many times the algorithm under discussion
produces a result that is within $0.000001$ of the optimal result (among $10$ trials).

\begin{footnotesize}
\begin{table}
\begin{center}
    \begin{tabular}{ |l|l|l|l| }
 \hline
        & Success Rate & Time[s] & \#Opt \\
 \hline
        {\sc BudgetFPT} & $0.308$ & $5.517$ &  \\
 \hline
        {\sc GCN} & $0.324$ & $34.166$ & $9$ \\
 \hline
        {\sc Greedy} & $0.483$ & $0.117$ & $1$ \\
 \hline
\end{tabular}
    \caption{\label{tab:experiment1}{\sc r2000} with $b=10$ and $s=5$}
\end{center}
\end{table}
\end{footnotesize}

\noindent {\bf Table~\ref{tab:experiment2}: Acyclic setting.}
We use {\sc r2000-dag} instead.
We compare {\sc DP} against {\sc Greedy} and {\sc GCN}.
It turns out that both {\sc GCN} and {\sc Greedy} always achieve the optimal
results. {\sc Greedy} being optimal is not entirely surprising, because we have
shown that {\sc Greedy} is optimal if the graph is exactly a tree, and
{\sc r2000-dag} is very close to a tree.

\begin{footnotesize}
\begin{table}
\begin{center}
    \begin{tabular}{ |l|l|l|l| }
 \hline
        & Success Rate & Time[s] & \#Opt \\
 \hline
        {\sc DP} & $0.449$ & $0.007$ &  \\
 \hline
        {\sc GCN} & $0.449$ & $2.340$ & $10$\\
 \hline
        {\sc Greedy} & $0.449$ & $0.017$ & $10$ \\
 \hline
\end{tabular}
    \caption{\label{tab:experiment2}{\sc r2000-dag} with $b=10$ and $s=5$}
\end{center}
\end{table}
\end{footnotesize}

\noindent {\bf Table~\ref{tab:experiment3}: Acyclic setting with substitutable edges.}
We could artificially modify {\sc r2000-dag} and create an acyclic graph that
is not close to a tree. We recall that {\sc Greedy} fails when there are
substitutable blockable-worthy edges, as shown in Figure~\ref{fig:example}.
We introduce substitutable edges into {\sc r2000-dag}.  Given a
blockable edge $a\rightarrow b$, we duplicate $b$.  We
create two paths $a\xrightarrow{blockable} b\rightarrow c$ and
$a\xrightarrow{blockable} b'\rightarrow c$, where $c$ is a successor of $b$.
Essentially, we are recreating the structure of Figure~\ref{fig:example}. This
setup is actually simulating a practical scenario. If there are two different
exploits that allow the attacker to travel from $a$ to $b$, then we need to
block both exploits. Under this artificially created acyclic graph, {\sc GCN}
is still always getting the optimal result, and {\sc Greedy} fails $4$ out of $10$
trials.



\begin{footnotesize}
\begin{table}
\begin{center}
    \begin{tabular}{ |l|l|l|l| }
 \hline
        & Success Rate & Time[s] & \#Opt\\
 \hline
        {\sc DP} & $0.465$ & $0.008$ &  \\
 \hline
        {\sc GCN} & $0.465$ & $8.000$ & $10$\\
 \hline
        {\sc Greedy} & $0.566$ & $0.030$ & $6$ \\
 \hline
\end{tabular}
    \caption{\label{tab:experiment3}{\sc r2000-dag} with additional substitutable edges, $b=10$ and $s=5$}
\end{center}
\end{table}
\end{footnotesize}

\noindent {\bf Table~\ref{tab:experiment4}: Setting without optimal solutions.}
We go back to {\sc r2000}. We double the budget to $20$ and double the number
of entry nodes to $10$. We can no longer afford to run {\sc BudgetFPT}.  We
also cannot run {\sc DP} because {\sc r2000} contains cycles.  In this setting,
we do not have optimal solutions.  We compare {\sc Greedy} against {\sc GCN}.
\#Win refers to how many times an algorithm beats the other one (ties are not
counted as wins).
It turns out that {\sc Greedy} performs better than {\sc GCN} in this setting.
In {\sc r2000}, there are naturally occurring substitutable edges.
With double budget, {\sc Greedy} will block the substitutable edges despite they are perceived as low in priority ({\em i.e.} no place to spend the budget, so might as well
spend on substitutable edges).
If we double $p_b$, then {\sc Greedy} again has plenty of places to spend budget on, so it will not spend budget on substitutable
edges. As a result, {\sc GCN} constantly wins again.

\begin{footnotesize}
\begin{table}
\begin{center}
    \begin{tabular}{ |l|l|l|l| }
 \hline
        & Success Rate & Time[s] & \#Win \\
 \hline
        {\sc GCN ($p_b=0.2$)} & $0.450$ & $37.987$ & $1$\\
 \hline
        {\sc Greedy ($p_b=0.2$)} & $0.438$ & $0.245$ & $3$\\
 \hline
 \hline
        {\sc GCN ($p_b=0.4$)} & $0.226$ & $24.121$ & $5$\\
 \hline
        {\sc Greedy ($p_b=0.4$)} & $0.287$ & $0.600$ & $0$\\
 \hline
\end{tabular}
    \caption{\label{tab:experiment4}{\sc r2000} with $b=20$ and $s=10$}
\end{center}
\end{table}
\end{footnotesize}

\section{Conclusion}

We studied edge blocking for defending Active Directory
style attack graphs. We proposed $3$ fixed-parameter algorithms based on the
observation that practical Active Directory attack graphs have small maximum
attack path lengths and are similar to trees.
{\sc BudgetFPT} can be applied when both the budget and the number of entry nodes
are small.
{\sc DP} scales the best experimentally, but it is only applicable to acyclic graphs.
{\sc SplitFPT} is based on performing fixed-parameter analysis on the attacker's strategy space, which
happens to be much smaller compared to the defensive strategy space in our model.
For each attacking strategy, we analysed what kind of defense would make this attacking strategy a
        valid best response. This FPT technique is potentially useful for other
        Stackelberg games.
Lastly, we {\em scaled up} {\sc SplitFPT}
by converting it to a graph convolutional neural network based heuristic.
A typical FPT approach is to exhaustively search among a set of solutions
characterized by the fixed parameters. Instead of exhaustive search, we trained
a neural network to guess the best solution.
Our algorithms will help IT admins
to identify high-risk edges (accesses/exploits) in practical Active Directory
environments.


\section*{Acknowledgements}
Frank Neumann has been supported by the Australian Research Council through grant FT200100536.
Hung Nguyen is partially supported by the ``Cyber NGT – Provable Network Security'' grant.

\bibliographystyle{unsrtnat}
\bibliography{/home/mingyu/Dropbox/nixos/mg.bib,/home/mingyu/Dropbox/nixos/mggame.bib}

\section*{Appendix}

\subsection{Proof of Theorem 1}

\begin{proof} We reduce {\em clique} to our problem.
    Clique is known to be $w[1]$-hard with respect to the clique
    size~\cite{Fellows2009:parameterized}.
    Let $G'$ be an arbitrary clique instance.
    $G'$ is an undirected graph.
    For each edge $e$ from $G'$, we construct one corresponding entry node $e_s$ in attack graph $G$.
    For every node $i$ from
    $G'$, we construct two nodes $i_{in}$ and $i_{out}$ in attack graph $G$.
    We assume $e$ connects node $i$ and node $j$ from $G'$.  We create two paths
    $e_s\rightarrow i_{in}\xrightarrow{blockable} i_{out}\rightarrow {\sc DA}$ and
    $e_s\rightarrow j_{in}\xrightarrow{blockable} j_{out}\rightarrow {\sc DA}$.
    That is, $e_s$'s shortest path to {\sc DA} has length $3$ unless both
    $i_{in}\rightarrow i_{out}$
    and $j_{in}\rightarrow j_{out}$ are blocked.
    The best way to spend a defensive budget of $b$ is to locate a clique
    of size $b$ in $G'$. Only by locating a clique, the defender can disconnect
    a maximum of $b(b-1)/2$ entry nodes.

    We then calculate the size of the constructed attack graph $G$.
    If $G'$ has $n'$ nodes and $m'$ edges, then the attack graph $G$ has
    $m'+2n'+1$ nodes (one node for each edge in $G'$, two nodes for each node
    in $G'$, and finally {\sc DA}) and $2m'+2n'$ edges (two edges from each
    $e_s$ and for each node in $G'$, one edge for connecting $i_{in}$ and
    $i_{out}$ and one edge for connecting $i_{out}$ to {\sc DA}).
    If a fixed-parameter tractable algorithm exists with respect to $b$ for the attack graph $G$, then its complexity must be $O(f(b)poly(m'+2n'+1))$,
    which is then $O(f(b)poly(n'))$. But then this means it is fixed-parameter
    tractable to find a clique of size $b$ in the Clique instance $G'$, which
    is contradictory to Clique being $w[1]$-hard.
\end{proof}

\begin{algorithm}[h!]
    \caption*{{\sc BudgetFPT}($G$, $b$)}\\
\textbf{Input}: Attack graph $G$, budget $b$\\
\textbf{Output}: Attacker's expected success rate\\
\begin{algorithmic}[1]
\IF {$b$ equals $0$}
    \STATE Breadth-first search from {\sc DA} and record all entry nodes'
    distances to {\sc DA}
    \FOR {every entry node $s_i$}
    \STATE Add $f(dist(s_i))/s$ to expected success rate
    \ENDFOR
    \STATE Return expected success rate
\ENDIF
    \IF {no entry nodes left}
    \STATE Return $0$
\ENDIF
\STATE Pick an arbitrary entry node $s_1$
\STATE Compute an arbitrary shortest path $p$ from $s_1$ to {\sc DA}
\STATE Get $G'$ by labelling $s_1$ as an non-entry node
    \STATE Record {\sc BudgetFPT}($G'$, $b$) plus $f(dist(s_1))/s$
\STATE Revert back to $G$
\FOR {blockable edge $e$ in $p$}
\STATE Get $G'$ by removing $e$
\STATE Record {\sc BudgetFPT}($G'$, $b-1$)
\STATE Revert back to $G$
\ENDFOR
\STATE Return the minimum recorded value
\end{algorithmic}
\end{algorithm}

\subsection{Proof of Proposition 1}

\begin{proof}
We have the following recursive relationship and the base cases:
\[c_{b,s} = lc_{b-1,s} + c_{b,s-1}\]
\[c_{0,s} = 1, c_{b,1} = l^b \]




We define the $m_{i,j}$ as follows
\[m_{i,j} = c_{i,j+1}/l^i\]

As a result, we have the following much more elegant recurrence relationship:
\[m_{i,j} = \frac{lc_{i-1,j+1} + c_{i,j}}{l^i}= m_{i-1,j}+m_{i,j-1}\]
\[m_{0,j} = 1, m_{i,0} = 1 \]

We use the standard generating function trick to analyse the recurrence
    relationship for $m_{i,j}$. We define $f(x,y)$ as follows

\[f(x,y)=\sum_{i,j\ge 0}m_{i,j}x^iy^j\]

\[f(x,y)=\sum_{i,j\ge 0}m_{i,j}x^iy^j=\sum_{i\ge 1,j\ge 1}m_{i,j}x^iy^j\]
\[
+\sum_{i\ge 1,j=0}m_{i,j}x^iy^j
+\sum_{i=0,j\ge 1}m_{i,j}x^iy^j+1
\]
\[=\sum_{i\ge 1,j\ge 1}m_{i,j}x^iy^j
+x\sum_{i\ge 0,j=0}m_{i,j}x^iy^j
+y\sum_{i=0,j\ge 0}m_{i,j}x^iy^j+1
\]
\[=\sum_{i\ge 1,j\ge 1}m_{i-1,j}x^iy^j+\sum_{i\ge 1,j\ge 1}m_{i,j-1}x^iy^j
\]
\[
+x\sum_{i\ge 0,j=0}m_{i,j}x^iy^j
+y\sum_{i=0,j\ge 0}m_{i,j}x^iy^j+1
\]
\[=x\sum_{i\ge 0,j\ge 1}m_{i,j}x^iy^j+y\sum_{i\ge 1,j\ge 0}m_{i,j}x^iy^j
\]
\[
+x\sum_{i\ge 0,j=0}m_{i,j}x^iy^j
+y\sum_{i=0,j\ge 0}m_{i,j}x^iy^j+1
\]
\[=(x+y)f(x,y)+1\]

    Since $f(x,y)=(x+y)f(x,y)+1$, we have
\[f(x,y)=\frac{1}{1-x-y}=1+(x+y)+(x+y)^2+(x+y)^3+...\]

$m_{i,j}$ is from the term $(x+y)^{i+j}$. The coefficient for $x^iy^j$ is $m_{i,j}={i+j\choose i}$. Therefore, we have

\[c_{b,s}= l^b m_{b,s-1} = l^b {b+s-1 \choose b}\]

Above we have calculated the number of combinations.
For each combination, performance evaluation takes linear time, so
the overall complexity is
    \[O\left(l^b {b+s-1 \choose b}n\right)\]

It should be noted that according to the pseudocode, we need to calculate
    shortest path (linear time) over and over again to generate all combinations.
When budget
is $1$, we calculate one shortest path (linear time) and this cost
    is distributed on $l$ combinations. When budget equals $2$, we
    calculate one shortest path and this cost is distributed on $l^2$ combinations.
    For each combination, the cost distributed on it for computing shortest
    path is linear time, multiplied by $\frac{1}{l}+\frac{1}{l^2}+\ldots \le 2$. Therefore,
    for each combination, the computation time is still linear.
\end{proof}

\subsection{Pseudocode for Dynamic Program}

\noindent {\bf Step 1: generate a desired tree decomposition}

\begin{algorithm}[h]
\caption*{Desired tree decomposition heuristic}
\begin{algorithmic}[1]
\STATE Topologically sort $G$ into $L$ security levels so that edges
only go from lower-level nodes to higher-level nodes\\
\FOR {security level $l$ from $1$ to $L$}
    \FOR {node $i$ in level $l$}
    \STATE Let $N(i)$ be $i$'s neighbours under current graph
    \STATE Create tree node $T_i = \{i\}\cup N(i)$
    \FOR {every pair of vertices $a,b\in T_i$}
        \STATE Add edge $(a,b)$ to $G$
    \ENDFOR
\ENDFOR
\ENDFOR
\FOR {$T_i$ from $1$ to $n$}
    \STATE Let $j=\max\{ T_i/\{i\}\}$
    \STATE Connect tree node $T_i$ to $T_j$
\ENDFOR
\end{algorithmic}
\end{algorithm}

\noindent {\bf Step 2: generate auxiliary tree nodes according to the {\em nice} tree decomposition idea (Cygan 2015)}

\begin{algorithm}[h]
    \caption*{Add auxiliary tree nodes}
\begin{algorithmic}[1]
    \STATE For every tree node $a$, clone itself and insert the clone $a'$ (labelled as an auxiliary node) in between
itself and its predecessors. For example, if the predecessors are $a_1,a_2,a_3$,
then remove $a_1,a_2,a_3$ from $a$ and attach them to $a'$ instead. Attach $a'$ to $a$.
    \WHILE {there exists node $a$ with more than two predecessors}
    \STATE For example, $a$ has three predecessors $a_1,a_2,a_3$. Clone $a$ into $a'$ (labelled as an auxiliary node). Remove $a_2,a_3$ from $a$. Attached $a_2,a_3$ to $a'$ instead. Attached $a'$ as $a$'s second predecessor.
    \ENDWHILE
\end{algorithmic}
\end{algorithm}

\newpage
\noindent {\bf Step 3: dynamic program subproblems}

\begin{itemize}

    \item Subproblem on {\em non-auxiliary} node $(x,y_1,y_2,\ldots)$ and remaining budget $b'$:
        The task is to optimally allocate budget on this node (remaining budget is passed down to current node's only predecessor, which must be an auxiliary node), in order to minimize the expected success rate of the attacker assuming that the attacker starts from an entry node that belongs to the current subtree.
        The context information is the $y_i$s' distances to {\sc DA}.
        We decide how many units of budget to spend on $x$ itself to block its
        out-going edges to the $y_i$. If $y_i$'s distance to {\sc DA} is $dist(y_i)$,
        then unless the edge $x\rightarrow y_i$ is blocked (or it does not exist), $x$ has a path of length
        $dist(y_i)+1$ to {\sc DA}.
        If $z$ units is spent on $x$, then we will always be blocking
        the shortest $z$ options.\footnote{We need to sort $w$ options. We treat this as $O(1)$ in complexity analysis, because our algorithm can only realistically handle single digit tree width $w$. Alternatively, we could multiply $wlog(w)$ to the complexity result.} We go through $b'+1$ options (spend $0$ to $b'$) to find the optimal unit of budget to spend on $x$ itself.
        Once $x$'s distance is decided, the distance is passed down as part of the
        context information to the only predecessor node, which must be an auxiliary node.
        The remaining budget is also passed down. The result of the current subproblem is equal to the result of the predecessor's subproblem. If $x$ is an entry node, we also need to add $f(dist(x))/s$ to the result of the current subproblem.

    \item Subproblem on {\em auxiliary} node $(y_1,y_2,\ldots)$ and remaining
        budget $b'$: The task is to optimally split the remaining budget to its
        predecessor nodes, in order to minimize the expected success rate of
        the attacker assuming that the attacker starts from an entry node that
        belongs to the current subtree.
        There are at most two predecessors. We need to divide
        $b'$ between the two predecessors.  That is, pass $b'_1$ to one
        predecessor and $b'_2=b'-b'_1$ to the other predecessor.  Sum up the
        two return values from the two predecessors' subproblems.  The
        splitting scheme corresponding to the minimum sum is the chosen scheme.

\end{itemize}


\begin{algorithm}[h]
    \caption*{{\sc SplitFPT}}
\begin{algorithmic}[1]
    \STATE An attacker strategy is a vector of size $t$ ($t$ is the number
    of splitting nodes). Coordinate $i$ specifies which route the attacker
    would choose at splitting node $i$ when facing the optimal defence.
    If splitting node $i$ has $d_i$ out-going edges, then coordinate $i$'s
    potential values include $0$ to $d_i-1$, as well as $-1$ (indicating the
    attacker does not choose any route).
    \FOR {every attacker strategy}
        \FOR {every splitting node $i$ where the attacker's strategy value is not
        $-1$}
        \STATE Let $j$ be the successor that the attacker chooses to travel to
        \STATE Follow along the simple path starting from $i\rightarrow j$ until
        we reach another splitting node or {\sc DA}. Label
        all edges along the way as not blockable
        \ENDFOR
        \STATE Calculate all splitting nodes' distances to {\sc DA}. This can be
        done by constructing the spanning tree formed by the splitting nodes and {\sc DA}
        \FOR {every route (successors of splitting nodes) not taken by the attacker}
        \STATE Follow along the simple path starting from this route until
        we reach another splitting node or {\sc DA}
        \IF {the simple path is not blocked
        and this route is shorter
        than the attacker's chosen route}
        \STATE Block the blockable edge that is closest to {\sc DA} along this simple path. If blockable edges do not exist, flag the current strategy as invalid.
        \ENDIF
        \ENDFOR
        \STATE Run {\sc Greedy on exact trees} using the remaing budget, record the
        minimum expected success rate returned.
    \ENDFOR
\end{algorithmic}
\end{algorithm}

\begin{algorithm}[h]
    \caption*{{\sc Greedy on exact trees} (subroutine of {\sc SplitFPT})}\\
\textbf{Input}: Remaining budget $b'$\\
\begin{algorithmic}[1]
    \STATE Breadth-first search and label all {\em frontier} blockable edges.
    An edge $e$ is a frontier blockable edge if there are no blockable edges
    along the path from $e$ to {\sc DA}.
    \FOR {each frontier blockable edge $e$}
    \STATE Calculate the benefit of removing $e$ via a breadth-first search to locate
    all entry nodes that can be cut off by blocking $e$. (The whole for loop takes
    linear time.)
    \ENDFOR
    \STATE Use Fibonacci heap to pick the most beneficial $b'$ edges to block.
    $O(1)$ for each insertion (at most $O(n)$ insertions) and $O(log(n))$ for each deletion (at most $b'$ deletions). Combined is in $O(b'n)$.
    \STATE Return the attacker's expected success rate after blocking the most
    beneficial $b'$ edges
\end{algorithmic}
\end{algorithm}

\subsection{Proof of Proposition 2}

\begin{proof}
There are $O(n)$ nodes. For each node, there are $w+1$ coordinates for tracking distances.  Each coordinate has $l+2$ options ($0$ to $l$, and infinity).
Every subproblem also tracks the remaining budget.
    The number of subproblems is $O((l+2)^{w+1}nb)$.
    Finally, each subproblem is an aggregation of at most $b+1$ subproblems.
\end{proof}

\subsection{Proof of Proposition 3}

\begin{proof}
The number of strategy combinations for the attacker is $\prod_{i=1}^{t}(d_i+1)$,
where $t$ is the number of splitting nodes and $d_i$ is the out degree of splitting
node $i$.
This value is at most $3^h$, because if there are $h$ feedback edges, spreading
    them on $h$ different splitting nodes (out degree $2$) would maximize the number of combinations for the attacker's strategies.

Given a specific attacker strategy, we have to mark all taken simple paths as
unblockable. There are at most $2h$ simple paths. The length is at most $l$
    each. So this step takes $O(hl)$.
    We also need to construct a spanning tree for the splitting nodes in order
    to calculate these nodes' distances to {\sc DA} under the current attacking
    strategy. We could implement a small graph with splitting nodes only,
    and the edges are weighted (based on the lengths of the corresponding simple paths).
    The number of splitting nodes is at most $h$.
    Assuming a simple adjacency matrix implementation, this step takes $O(h^2)$.
    We then need to check and block all simple paths not taken.
    This steps takes $O(hl)$.
    Finally, we need to run {\sc Greedy} to spend the remaining budgets.
    {\sc Greedy} takes linear time $O(n)$ to locate one edge to block when the
    attack graph is a tree. It is repeated $b$ times at most. So this step
    takes $O(bn)$.
\end{proof}

\subsection{Node features used in graph convolutional neural networks}

\begin{itemize}

    \item A node's security level (based on topological sorting)

    \item In degree

    \item Out degree

    \item Number of entry nodes that can reach this node

    \item Distance to {\sc DA} assuming no edges are blocked

    \item Distance to {\sc DA} assuming all edges are blocked

    \item Whether this node is an entry node
\end{itemize}

\subsection{Edge features used in graph convolutional neural networks}

Every edge in the condensed graph corresponds to a simple path from the original graph.

\begin{itemize}

    \item First node of the simple path

    \item Length of the simple path

    \item How many edges can be blocked along the simple path

    \item Security level of the destination node from the blockable edge
        that is closest to {\sc DA}

    \item How many entry nodes are along the simple path

    \item Direction (incoming or out-going)
\end{itemize}

\end{document}